\documentclass[prl,showpacs,twocolumn]{revtex4}

\usepackage{epsfig}
\usepackage{color}
\usepackage{mhchem}
\usepackage[makeroom]{cancel} 
\usepackage[normalem]{ulem} 
\usepackage{amssymb,graphics,graphicx,amsfonts,amsmath,empheq}
\usepackage{pdfpages}

\def\d{\mathrm{d}}
\def\e{\mathrm{e}}

\def\P{\mathcal{P}}

\begin{document}
\title{Ciliary beating patterns map onto a low-dimensional behavioral space that accords with a simple mechanochemical model}


\author{Veikko F. Geyer}
\affiliation{B CUBE--Center for Molecular Bioengineering, Dresden, Germany}

\author{Jonathon Howard}
\affiliation{Department of Molecular Biophysics and Biochemistry, Yale University, New Haven, United States}

\author{Pablo Sartori}
\affiliation{Instituto Gulbenkian de Ciencia, Oeiras, Portugal}

\date{\today}

\begin{abstract}
Biological systems are robust to  perturbations at both the genetic and environmental levels. Yet, these same perturbations can elicit variation in behavior. The interplay between functional robustness and behavioral variability is exemplified at the organellar level by the beating of cilia and flagella. The complex bending patterns of cilia emerge from the coordinated activities of hundreds of different proteins. Cilia are motile despite wide genetic diversity between and within species, large differences in intracellular concentrations of ATP and calcium, and considerable environment fluctuations in temperature and viscosity. At the same time, these perturbations result in a variety of spatio-temporal patterns that span a rich, and as of yet uncharted, behavioral space. To investigate this behavioral space we analyzed the dynamics of isolated cilia from the unicellular algae {\it Chlamydomonas reinhardtii} in thirty four different environmental and genetic conditions. We found that, despite large changes in beat frequency and amplitude, the space of waveform shapes is low-dimensional. Two features, parabolicity and asymmetry of the tangent angle, account for eighty per cent of the  observed variation. Remarkably, the geometry of this behavioral space accords with the predictions of a simple mechanochemical model in the low viscosity regime. This allowed us to associate waveform shape variability with changes in only the curvature response coefficients of the dynein motors. Our work demonstrates that, despite the molecular complexity of the cilium, the space of beat variations is low dimensional and can be attributed to variation in mechanochemical regulation of the dynein motors.
\end{abstract}

\maketitle



\section{Introduction}

Biological systems can function despite genetic and environmental perturbations in their molecular components. These perturbations, in turn, affect behavior at higher levels, from organelles through cells to whole organisms. A key finding is that behavioral spaces, the spaces that embed the whole repertoire of behavior, tend to be low dimensional. Examples are the shapes of moving nematodes  \cite{stephens2008dimensionality, helms2019modelling}, the swimming trajectories of ciliates and bacteria \cite{jordan2013behavioral, plevska2021nongenetic}, the postures of walking flies \cite{berman2016predictability} and mice \cite{mathis2018deeplabcut}, and the activities of Hydra \cite{han2018comprehensive}, which can all be reduced to low-dimensional behavioral spaces. An important conclusion of these analyses is that only a few features define each behavior, though there can be  considerable variation of these features among individuals within that behavior. A challenge is to map these behavioral spaces onto underlying molecular and biophysical mechanisms \cite{wang2020neuron}.

Cilia are motile organelles that undergo complex shape changes. These changes drive motion relative to the surrounding fluid and allow cilia to respond to external cues. Cilia, therefore, display a simple form of behavior that makes them a potentially good model system for relating behavior to underlying molecular mechanisms.  A key feature of cilia is that they can beat in the presence of a wide range of genetic mutations of key proteins \cite{BrokawKamiya1987, BrokawLuck1985, wirshell2013}, and can operate over a wide range of  temperatures \cite{Rikmenspoel1984}, viscosities \cite{Rikmenspoel1984,Yagi2005}, ATP concentrations \cite{Brokaw1975}, and buffer conditions like pH and calcium concentration \cite{BESSEN1980, Omoto1985, Ishijima1987}. Thus, cilia exemplify conserved function \cite{wan2020unity}, i.e. beating,  in the face of large genetic and environmental perturbations. In this work we use variations in beat shapes elicited by these perturbations to construct the behavioral space of cilia.

The beating motion of cilia is powered by thousands of dynein motors belonging to several different classes \cite{HuiBui2012}. These motors drive sliding of adjacent doublet microtubules \cite{summers1971adenosine} within the axoneme, the structural core of cilia \cite{nicastro2006molecular}. Sliding is converted to bending by proteins that constrain shearing of doublets \cite{Summers1971, Warner1976, brokaw2009thinking, Heuser2009} and sliding at the base \cite{satir1965studies}. This biomechanical and structural knowledge forms the basis of mechanochemical models that can successfully recapitulate the oscillatory motion of cilia \cite{machin1958wave, brokaw1985computer, lindemann1994geometric, bayly2016steady, bayly2014equations, riedel2007molecular, sartori2016dynamic} and synthetic  filament bundles \cite{sanchez2012spontaneous, pochitaloff2018self}. The combination of mechanistic models, described in this paragraph, with phenotypic variability in response to perturbations, described in the previous paragraph, makes the cilium an ideal model system to study the connection between behavioral spaces and molecular mechanisms.

In this work, we have asked: what is the geometry of the behavioral space of beating cilia subject to a wide range of  perturbations, and how do the individual components of this space relate to the underlying mechanochemistry.

\section{Results}

\begin{figure}
\includegraphics{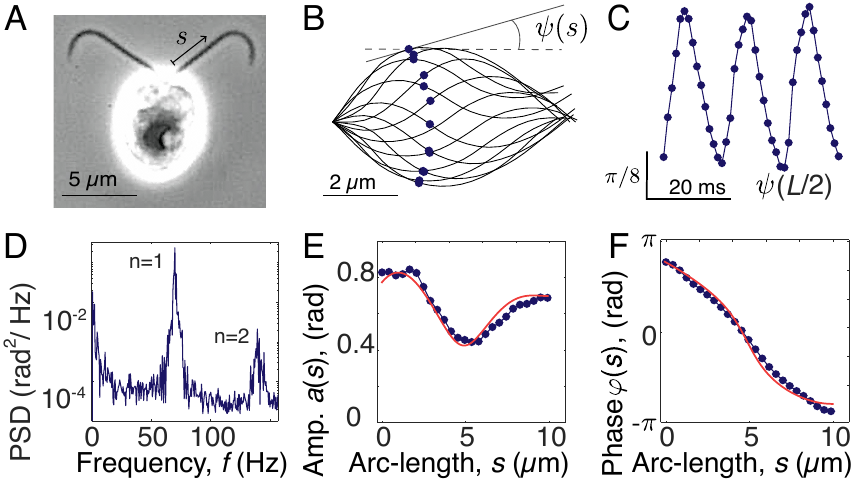}
\caption{\label{fig:scheme}
 {\it Quantification of ciliary  beats.} {\bf A} Wild-type {\it Chlamydomonas} cell imaged by phase-contrast microscopy. {\bf B} Tracked waveforms for one beat cycle in the co-swimming frame ($1\,{\rm ms}$ between curves) after the static mode was removed. Blue dots denote midpoint ($s=L/2$). {\bf C} Tangent angle at the midpoint as a function of time. {\bf D} Power spectral density of the tangent angle averaged over arc-length. {\bf E} Amplitude of the fundamental mode ($n=1$) as a function of arc-length. The dip in the middle is characteristic of wild-type axonemes. {\bf F} Phase of the fundamental mode as a function of arc-length. The approximately linear decrease in phase indicates steady propagation of the wave from the base (cell body). The red  curves in E and F correspond to the mechanochemical model described in the {\it Materials and Methods}.
}
\end{figure}

{\it Quantification of the ciliary beat.} We measured the waveforms of isolated and reactivated {\it Chlamydomonas reinhardtii} axonemes (Figure~\ref{fig:scheme}A) with high temporal and spatial precision, see \cite{sartori2016dynamic} and {\it Materials and Methods}. The waveforms, discretized in 25 points along the arc-length of the axoneme, were tracked over time for up to 200 beat cycles using filament tracking software \cite{ruhnow2011tracking}  (one beat cycle is shown in Figure~\ref{fig:scheme}B).

The periodic beat of the axoneme was parameterized by the tangent angle $\psi(s,t)$ (with $s$ the arc-length and $t$ time) relative to the co-swimming frame \cite{geyer2018computational}, Figure~\ref{fig:scheme}D. The power spectrum of the tangent angle (Figure~\ref{fig:scheme}E) showed that $<10\%$ of the power was in the modes $n\ge2$, which we therefore neglected. After subtracting the static mode $n=0$, discussed in  \cite{geyer2016independent, sartori2016dynamic}, we parametrized the beat dynamics as a traveling wave
\begin{align}\label{eq:shape}
\psi(s,t)=a(s)\cos(2\pi f t + \varphi(s))\quad,
\end{align}
where $a(s)$ is the amplitude profile, $\varphi(s)$ is the phase profile, and $f$ is the beat frequency. The arc-length was normalized by the total length $L$ of the axoneme, so that $s\in[0,1]$.
Although $a(s)$ and $\varphi(s)$ are often approximated by scalars such as the mean amplitude $a_0\equiv \int_0^1 a(s)\d s$ and the wavelength $\lambda\equiv -\frac{\pi L}{6}\int_0^1 (s-1/2)\varphi(s)\d s$, here we are interested in their arc-length dependence, which defines the shape of the beat. Our goal is to characterize variations in the set of parameters  $\P=\{a(s), \varphi(s), f, L\}$, which we define as the behavioral space of beating. A subset of these parameters, $\{a(s)/a_0, \varphi(s)\}$, defines the shape.

\begin{figure*}
\includegraphics{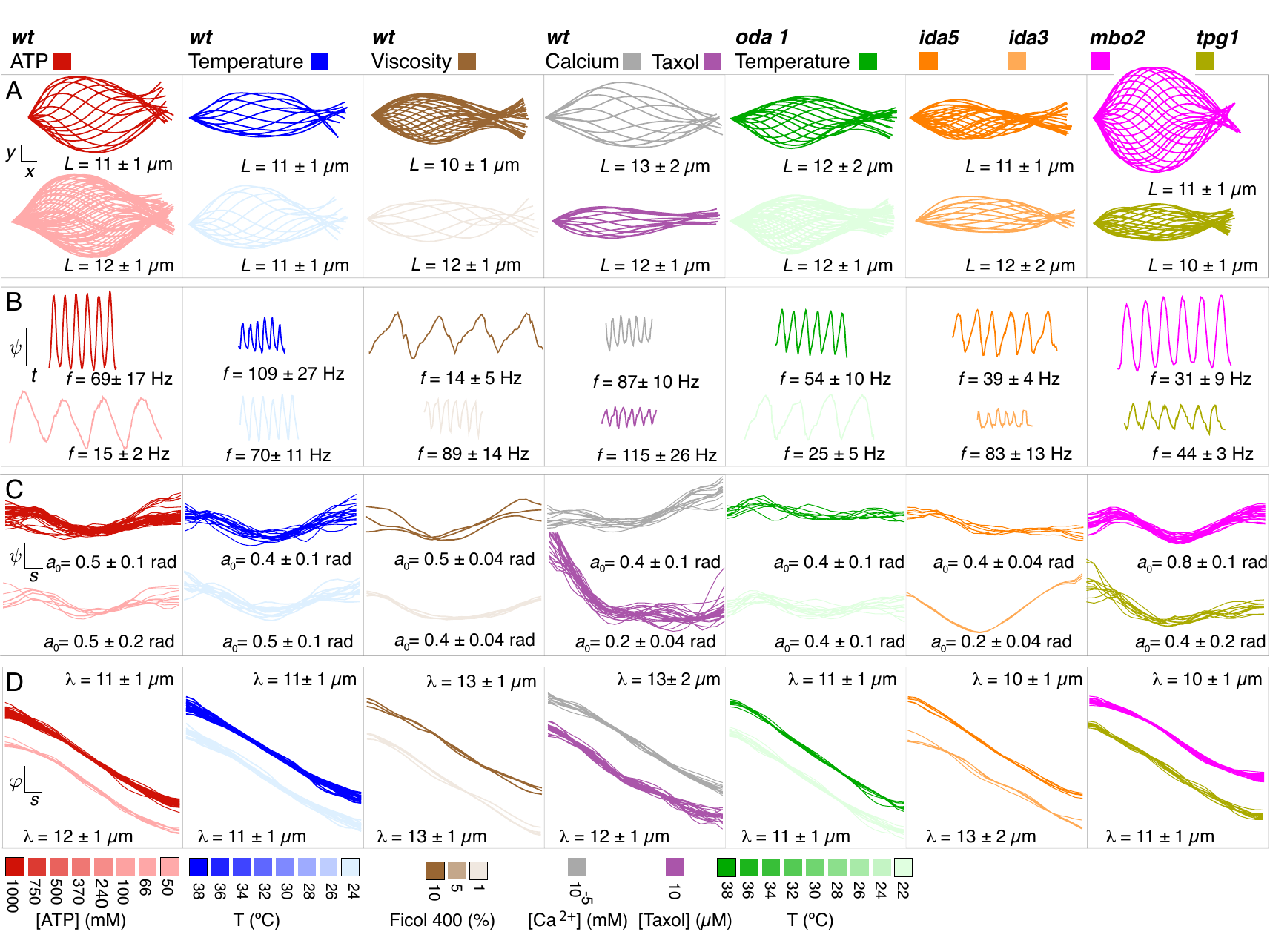}
\caption{
 {\it Behavioral variability of the ciliary beat under genetic and environmental perturbations.} Each column contains waveform data of a particular cell type and/or environmental condition (color legend below). Unless stated otherwise, reactivation was performed under standard conditions (24$^{\circ}$C and 1mM ATP, see {\it Materials \& Methods}). Means and standard deviations of beat parameters are provided in each row. {\bf A} Sequence of shapes for one beat cycle ($1\,{\rm ms}$ between curves, scale-bar $1\,\mu{\rm m}$). {\bf B} Tangent angle at the cilium midpoint as a function of time (scale-bars are  $20\,{\rm ms}$ and  $0.5\,{\rm rad}$). {\bf C} Amplitude profile $a(s)$, see Eq.~\ref{eq:shape} (scale-bars are $0.1  L$ and $0.5\,{\rm rad}$). {\bf D} Phase profile $\varphi(s)$, see Eq.~\ref{eq:shape} (scale-bars are $0.1 L$ and  $\pi/2\,{\rm rad}$).
 }\label{fig:table}
\end{figure*}

We explored the variability of $\P$ by observing groups of axonemes under several environmental (temperature, viscosity, different ATP and Ca$^{2+}$ concentrations, taxol) and genetic ({\it oda1}, {\it ida3}, {\it ida5}, {\it mbo2} and {\it tpg1}) perturbations, see Figure~\ref{fig:table}. Our dataset comprises 498 axonemes undergoing a total of  $\sim 40,000$ beat cycles. Under all conditions we found beating axonemes, with the percentage of reactivating axonemes greater than 70\%. The range of conditions resulted in a wide range of beat shapes and lengths (Figure~\ref{fig:table}A), beat frequencies (Figure~\ref{fig:table}B), amplitude profiles (Figure~\ref{fig:table}C), and phase profiles (Figue~\ref{fig:table}D). Therefore, while the beating of cilia is robust against perturbations,  the features of its waveform display substantial variations. We now set out to quantify these variations.

\begin{figure}
\includegraphics{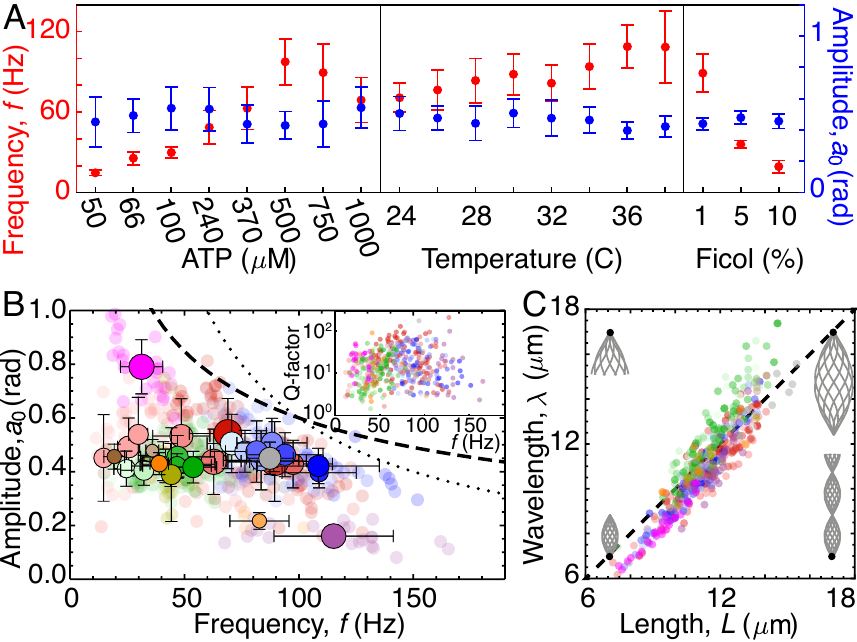}
\caption{{\it Variation in frequency, mean amplitude, length and wavelength.} {\bf A}. Effect on the mean amplitude and beat  frequency of changes in ATP concentration (left), temperature (middle), and viscosity (right). The frequency is strongly dependent on these environmental changes while the amplitude is not. {\bf B}. Scatter plot of amplitude versus frequency. Each transparent circle corresponds to one axoneme, and the solid circle indicates with the radius the size of the dataset. Both the mean amplitude and frequency vary by one order of magnitude. Lack of data with high amplitude and frequency may be due to energetic constraints. Dotted and dashed lines correspond to $1/f^{1/2}$ and $1/f$ scalings, respectively. Inset: scatter plot of Q-factor. {\bf C}. Scatter plot of wavelength vs length shows that most of the variability in wavelength is correlated to changes in length. Insets show waveforms with lengths/wavelengths corresponding to the black dots at their base.
}
\label{fig:amp_freq}
\end{figure}

{\it  Variations in frequency, mean amplitude and wavelength.}  The beat frequency increases with ATP concentration and temperature, and decreases with viscosity, Fig.~\ref{fig:amp_freq}A. This variation encompasses a frequency range of about one order of magnitude and the between-conditions variance is about seven times larger than the within-condition variance. By contrast, the mean amplitude varies little between these conditions, although it shows large variations within conditions: the between-conditions variance in amplitude is about one fifth the within-conditions variance. Thus, ATP concentration, temperature and viscosity lead to large changes in frequency with little effect on mean amplitude, in agreement with studies cited in the {\it Introduction}.

The mutations {\it oda1} (lacking outer-arm dyneins, \cite{Mitchell1985}), {\it ida3} and {\it ida5} (lacking subsets of inner-arm dyneins, \cite{Kamiya1991,Kato-Minoura1997}), {\it mbo2} (which lacks microtubule inner proteins and has a symmetric beat, \cite{Segal1984}) and {\it tpg1} (which has reduced polyglutamamylation required for axonemal integrity, \cite{Kubo2010,Alford2016}), as well as addition of the ion Ca$^{2+}$ and the microtubule-poison taxol, did not appreciably broaden the range of beat frequencies (15 Hz to 160 Hz, Figure 3B). We note, however, that the {\it oda1} mutant has a two-fold lower beat frequency over all temperatures, as reported in earlier studies (see Introduction). {\it mbo2}  and taxol increased the range of mean amplitudes (Figure 3B). Interestingly, in {\it mbo2} the amplitude is high but the beat frequency is low, while in taxol (which has not been studied before) the amplitude is low but the beat frequency is high; this reciprocal variation of amplitude and frequency may reflect the energetics of the beat (see {\it Discussion}). 

The variations in the beat frequency and mean amplitude within individual axonemes are small compared to the axoneme-to-axoneme variation. Over the times of analysis, which were typically $\sim 50$ cycles, the variances of the frequency and the mean amplitude were considerably smaller than the axoneme-to-axoneme variance, see Supplementary Table. Fluctuations in beat frequency over time are so small that the Q-factor of the oscillations can reach values as high as 150 (Figure 3B Inset); this Q-factor is one of the highest in any biological system, being exceeded by the tuning of sonar responses in the inner ear of the mustached bat (Q up to 1000, \cite{kossl1995basilar}). Thus, the variation in beat frequency and amplitude is not due to short-term variation within individual axonemes, but rather differences between axonemes and their response to perturbations.

The lengths of axonemes vary two-fold over the entire dataset, from 7.2 to 15.4 $\mu{\rm m}$, see Fig.~\ref{fig:amp_freq}C. This variation is primarily due to the shorter {\it mbo2} axonemes and the longer  {\it oda1} axonemes. Remarkably, the wavelengths under all these conditions are almost equal to the lengths. In other words, most of the variability in wavelength is explained by variability in length, such that the wavenumber is close to unity, $\lambda/L=0.97\pm 0.08$ (mean $\pm$ standard deviation for all 498 cilia). 
This finding is reminiscent, though different, to the standing waves in a string or pipe, and has important implications for the beat-generation mechanism (see analysis around Fig.~\ref{fig:model}).

\begin{figure}
\includegraphics{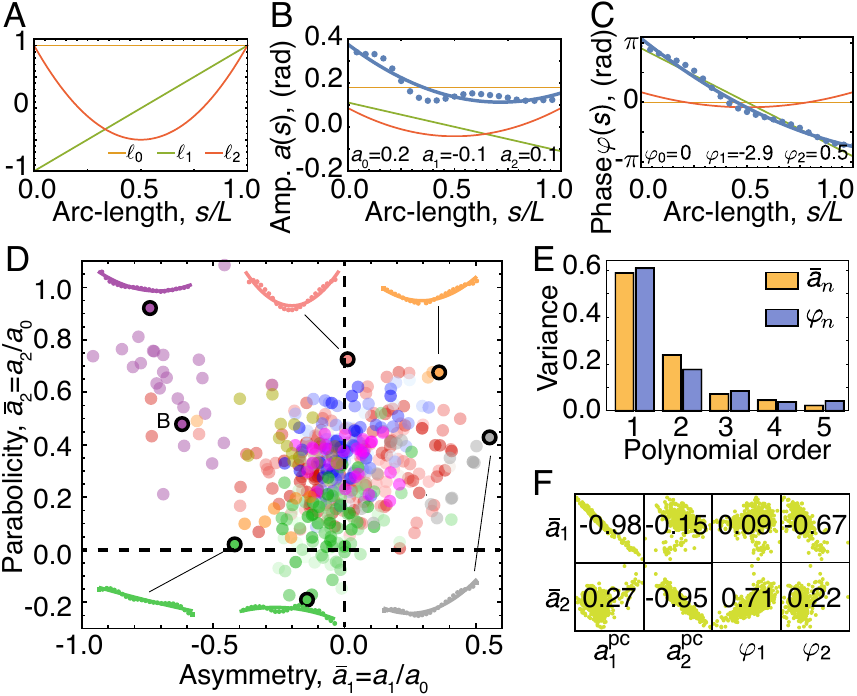}
\caption{{\it Decomposition of waveform variability.} {\bf A}. Shifted Legendre polynomials $\ell_n(s)$ for $n=0,1,2$. {\bf B}. Decomposition of a particular amplitude profile into the first three shifted Legendre polynomials. $a_0$ corresponds to the mean amplitude of Fig.~\ref{fig:amp_freq}, while $a_1<0$ and $a_2>0$ denote decreasing and convex amplitude profiles, respectively. {\bf C}. Same as B but for a phase profile. {\bf D}. Scatter plot of the normalized first and second polynomial coefficients for the amplitude. Insets show examples of amplitude profiles for different values of these coefficients. {\bf E}. Fraction of the variance in amplitude and phase profile explained by shifted Legendre polynomials of increasing order. {\bf F}. Table of correlation coefficients of amplitude polynomial coefficients with amplitude principal components and phase polynomial coefficients. Note that $\bar{a}_{1}$ and $\bar{a}_{2}$ are strongly correlated to the first two principal components, ${a}_{1}^{\rm pc}$ and $\bar{a}^{\rm pc}_{2}$); and are anti-correlated and correlated with ${\varphi}_{2}$ and ${\varphi}_{1}$. Green background is scatter-plot of data.}
\label{fig:robust}
\end{figure}

\begin{figure}
\includegraphics{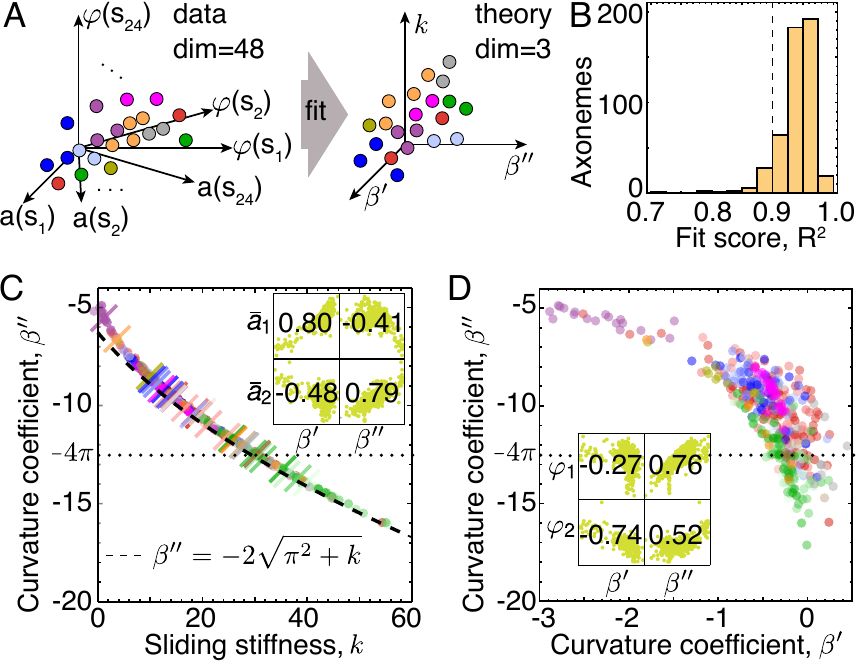}
\caption{
 {\it Mapping waveform-space to mechanical-space.} {\bf A}. Experimental waveforms have 48 dimensions, 24 for the amplitude and 24 for the phase profile. Fitting to a mechanical model, we map these data to a three dimensional mechanical space, with axis the two response coefficients of the motors and the shear stiffness. {\bf B}. Histogram of the $R^2$ of the fits. Over 90\% of the fits have $R^2>0.9$. {\bf C}. Variability in shear stiffness $k$ and motor response $\beta''$. The values of $k$ and $\beta''$ are tightly correlated, which corresponds to the low viscosity limit, dashed line. Dotted line corresponds to $-4\pi$.  The inset shows the correlation between curvature response coefficients and phase asymmetry and parabolicity. {\bf D}. Variability in curvature response coefficients captures waveform variability, with low negative $\beta'$ corresponding to low negative $\bar{a}_1$, and thus decaying waveforms. Dotted line corresponds to $-4\pi$. The inset shows the correlation between the curvature response coefficients and the asymmetry and parabolicity of the phase.}
\label{fig:model}
\end{figure}

{\it  Asymmetry and parabolicity are dominant waveform traits.} To characterize the variation in the amplitude and phase profiles of the tangent angle, we decomposed both into shifted Legendre polynomials of increasing order, see {\it Materials \& Methods} and Fig.~\ref{fig:robust}A. For the amplitude, the lowest order ($a_0$) equals the mean amplitude, the first order ($a_1$) corresponds to the linear deviation from constant amplitude and is a measure of the proximal-distal asymmetry, and the second order ($a_2$) measures the quadratic deviation. For the phase, we set $\varphi_0$ to zero (as it is arbitrary),  the first order is $\varphi_1=-\pi L/\lambda$  by definition of $\lambda$ (the linear component of the phase profile), and $\varphi_2$ corresponds to the quadratic deviation from the line. Because we are interested in shape, we normalized the amplitude coefficients: $\bar{a}_1=a_1/a_0$ and  $\bar{a}_2=a_2/a_0$.

Figure~\ref{fig:robust}D displays the space of variations of parabolicity and asymmetry of the amplitude. Most points cluster around $\bar{a}_1\approx0.0$ and $ \bar{a}_2\approx0.3$, which corresponds to a symmetric and convex amplitude profile. The exceptions include taxol, {\it ida5} and {\it tpg1}, which have $\bar{a}_1<0$ corresponding to decreasing amplitudes towards the distal tip, and Ca$^{\rm 2+}$, which has $\bar{a}_1>0$ corresponding to increasing amplitude towards the distal tip. {\it oda1} waveforms have a small parabolicity $\bar{a}_2$, corresponding to their linear amplitude profiles. {\it ida5} has smaller parabolicity and negative asymmetry. Thus, despite their robust beat, both genetic and environmental perturbations resulted in well-defined waveform differences.

Parabolicity and asymmetry are the two polynomial terms that contribute most to the variance, sixty and twenty percent respectively, see Fig.~\ref{fig:robust}E. Subsequent terms each contribute less than ten percent. Parabolicity and asymmetry are not only intuitive data features: they are the main features of the data, as revealed by their strong correlations with the first two principal components $a_1^{\rm pc}$ and  $a_2^{\rm pc}$, left boxes in Fig.~\ref{fig:robust}F. Furthermore, the asymmetry of the amplitude profile,  $\bar{a}_{1}$, is anti-correlated with the parabolicity of the phase profile,  ${\varphi}_{2}$, while the parabolicity of the amplitude profile, $\bar{a}_{2}$, is correlated with the asymmetry of the phase profile, ${\varphi}_{1}$, right boxes in Fig.~\ref{fig:robust}F.  Taken together, these facts demonstrate that, once the effects of mean amplitude and  length are taken into account, the dominant  shape features are the parabolicity and asymmetry of the amplitude and phase profiles.

{\it Motor response controls waveform variability.} Although we have studied a large number of different perturbations, the number of features is small. This suggests that perturbations, irrespective of their molecular origin, affect few collective properties of the axoneme.

To explore this possibility, we asked whether a mechanical model of the ciliary beat, based on that in \cite{sartori2016dynamic}, could account for the diversity of waveforms. The model, described in the {\it Materials \& Methods}, contains four non-dimensional coefficients: $\beta'$ and $\beta''$, which characterize the motors’ dependence on the instantaneous curvature (reactive response) and the rate of change of curvature (dissipative response) respectively; $k$, which is the shear stiffness of the axoneme; and $\overline{{\rm Ma}}$, which is a constant that comes from non-dimensionalizing Machin's equation, (see equation 4 \cite{machin1958wave}). $\overline{{\rm Ma}}={2\pi f\xi_{\rm n}L^4}/{\kappa}$, with $\kappa$ the flexural rigidity of the axoneme and $\xi_{\rm n}$ the friction coefficient of the medium. For {\it Chlamydomonas}, $\overline{{\rm Ma}}\sim50$, using the parameters in {\it Materials \& Methods}. In our earlier model \cite{sartori2016dynamic}, $\beta'= 0$ because only the symmetric wild-type and \textit{mbo2} axonemes were studied. We fitted the mechanical model to the data by determining $\overline{{\rm Ma}}$ for each axoneme and adjusting $\{\beta',\beta'',k\}$ to maximize the $R^2$ inverse distance between theoretical and experimental waveforms, see example in Fig.~\ref{fig:scheme}E,F. The curve fitting reduced the dimensionality of the waveform description from $48$, $a(s_i)$ and $\varphi(s_i)$ with $i=1,\ldots,24$ points along the length, to just $3$, the model parameters, see Fig.~\ref{fig:model}A. The agreement of the model with the data was excellent, with 92\% of the axonemes having $R^2> 0.9$, see Fig.~\ref{fig:model}B. 


Figure~\ref{fig:model}C shows that, despite there being three free fit parameters in the model, one of the parameters, the sliding stiffness $k$, is strongly correlated with another parameter,  the dissipative motor response $\beta''$. The observed relationship between these parameters can be recovered analytically by noting that the ratio of friction to bending forces, which is proportional to the Machin number ${{\rm Ma}}\equiv\overline{\rm Ma}/(2\pi)^4$, is $\ll1$. Taking the limit ${{\rm Ma}}\to0$ gives the curved dashed line in Fig.~\ref{fig:model}C, see {\it Materials \& Methods}. The good match between this ``dry friction'' limit and the full theory supports the hypothesis that {\it Chlamydomonas} operate in a low friction regime (see {\it Discussion}). Therefore, the waveform is controlled by just the two motor response coefficients, which correlate with the asymmetry and parabolicity of the amplitude, see  Fig.~\ref{fig:model}C inset. 

In agreement with the above argument,  Fig.~\ref{fig:model}D recapitulates the variability in waveform features observed in Fig.~\ref{fig:robust}D. For example, the taxol and Ca$^{2+}$ data lie far apart from each other in the $(\beta',\beta'')$ space, just as in the $(\bar{a}_1,\bar{a}_2)$ space.  Furthermore, because amplitude and phase are correlated,  the motor response coefficients also correlate with the phase features, see Fig.~\ref{fig:model}D inset. $\beta'$ and $\beta''$ not only control the asymmetry and parabolicity of the amplitude, respectively, but $\beta''$  also  controls the wavelength. Specifically, in the limit ${\rm Ma}\to0$, we can show that the curvature wavelength is $\lambda=-4\pi L/\beta''$, and so $\lambda/L=1$, which is typically observed, requires $\beta''=-4\pi$, in agreement with the average values in Fig.~\ref{fig:model}C (dotted line) and D. Thus, the space spanned by the motor-response coefficients accurately recapitulates the space of amplitude and phase features, which suggests that all genetic and environmental perturbations are buffered into changes of the curvature regulation of the motors.

\vspace{.1in}

\section{Discussion}
{\it The dimensionality of the ciliary beat is low.} By using a wide variety of environmental and genetic perturbations, we have constructed the behavioral space of ciliary swimming waveforms. Asymmetry and parabolicity of amplitude ($\bar{a}_1$ and $\varphi_1$) and phase ($\bar{a}_2$ and $\varphi_2$) profiles suffice to describe about 80\% of the variation between and within conditions. Pearson's correlation coefficient between $\bar{a}_1$ and $\varphi_2$ is $-0.67$ and between $\bar{a}_2$  and $\varphi_1$ is $0.71$. Therefore, just two parameters, $\bar{a}_1$ and $\bar{a}_2$, account for about 50\% of the phase variance (as well as about 80\% of the amplitude variance, Fig.~\ref{fig:robust}E). The dimensionality of swimming behavior is therefore very low: beat frequency, mean amplitude, shape asymmetry, shape parabolicity, and axoneme length. While low dimensional phenotypic spaces have been observed in other behaviors, see {\it Introduction}, the range of conditions probed in our study is unmatched, which makes the low dimensionality even more remarkable. 
\newline

 {\it The shape space is spanned by the curvature sensitivity of the dyneins.} The geometry of the shape space (the amplitude and phase profiles) accords with a simple biomechanical model of ciliary beat, which is an extension of a previous one \cite{sartori2016dynamic}. In this model, dynein motors respond to curvature with an instantaneous coefficient $\beta'$ and a dynamic coefficient  $\beta''$. The model also includes a passive stiffness to the sliding of doublets, $k$.

The connection between the shape and biomechanical spaces enabled a remarkable finding: the curvature coefficients $\beta'$ and $\beta''$ strongly correlate with the shape parameters, $\bar{a}_1$, $\bar{a}_2$, $\varphi_1$ and $\varphi_2$:
\begin{align}\label{eq:corr}
\bar{a}_1\sim\beta'\sim-\varphi_2\quad{\rm and}\quad\bar{a}_2\sim\beta''\sim+\varphi_1\propto\lambda/L\quad,
\end{align}
where $\sim$ denotes correlation. Thus, the different shapes correspond to different curvature sensitivities.

Wild-type cilia have symmetric and convex amplitude profiles, $\bar{a}_1\approx0.0$ and $ \bar{a}_2\approx0.3$, corresponding to $\beta'=0$ and $\beta''\approx-10$. Using Eq.~\ref{eq:corr}, we can ascribe deviations from this typical behavior to changes in motor properties. For example, taxol, {\it tpg1} and {\it ida5} have $\bar{a}_1<0$, corresponding to a beat whose amplitude decays distally. Despite the different nature of these perturbations, we find that all three perturbations decrease the instantaneous response coefficient, $\beta'$, and increase the dynamic response coefficient, $\beta''$. This means that the sensitivity to instantaneous curvature is increased, while the sensitivity to dynamic curvature is decreased. In other words, the phase of the curvature response is altered. Conversely, Ca$^{\rm 2+}$ leads to $\bar{a}_1>0$, and so $\beta'>0$. {\it oda1} has a flat profile, with $\bar{a}_2\approx0$ and $\bar{a}_1\approx0$, and so $\beta'\approx0$ and $\beta''\approx-12$. This counters the general belief that outer-arm dyneins only affect beat frequency \cite{porter20009}. {\it ida3} has a high parabolicity, i.e. a smaller amplitude in the middle, which reflects a smaller value of the dynamic curvature coefficient. {\it mbo2} has a waveform very similar to that of wild-type, which is remarkable given the absence of static mode in the waveform of this mutant \cite{geyer2016independent,segal1984mutant}, but consistent with the observation that the {\it mbo2} mutations do not affect dyneins \cite{PorterSale2000}. Thus, changes in the instantaneous and dynamic curvature sensitivity of the dyneins account for the effects of environmental and genetic perturbations.
\newline

{\it Elasticity dominates  {\it Chlamydomonas} swimming.} Our results contain three arguments that viscous forces are smaller than elastic forces for {\it Chlamydomonas} axonemes. First, the Machin number, which is the ratio of viscous to elastic forces and is related to the Weissenberg number \cite{poole2012deborah}, is much smaller than unity, ${\rm Ma}\ll1$. For a plane wave with wave number $q=2\pi/\lambda$ we have,
\begin{align}
{\rm Ma}=\frac{\xi_{\rm n} \omega}{\kappa q^4} = \frac{1}{(2\pi)^4}\overline{\rm Ma}\quad, 
\end{align}
which ranges from $0.02$ to $0.14$ in all conditions (note however that there is considerable uncertainty in $\kappa$, especially in the presence of taxol \cite{schaedel2015microtubules}). Second, the data collapse of the fitted $k$ and  $\beta''$ in Fig.~\ref{fig:model}C almost coincides with that predicted by the low-viscosity relation, dashed line. This extends our earlier finding that a low ${\rm Ma}$ accords with wild-type and {\it mbo2} axonemes \cite{sartori2016dynamic}. Third, despite the wide variability in frequency and mean amplitude, the total range is $\sim10$-fold for both, there is an absence of high-amplitude and high-frequency beats. The dissipation of elastic energy is proportional to $a_0^2f$ whereas the dissipation of viscous energy is proportional to $a_0^2f^2$ \cite{machin1958wave}. Thus, if elastic or viscous energy were limited by the energy available from the hydrolysis of ATP by the motors, then $a_0\sim f^{-1/2}$  or $a_0\sim f^{-1}$ respectively. The former provides a better bound on the data than the latter (Figure~\ref{fig:amp_freq}B dashed and doted curves, respectively), as expected for elastic dissipation dominating. 
	
The low viscosity regime is also supported by the literature. In \cite{sanchez2012spontaneous}, the ATPase rate of axonemes was shown to increase in proportion to beat frequency, as predicted for elastic dissipation, and in \cite{mondal2020internal}  measurements of the flow field around {\it Chlamydomonas} cilia showed that friction forces are smaller than the estimated bending forces. We conclude that {\it Chlamydomonas} swims in the regime of low ${\rm Ma}$, i.e. low friction. Because the Reynolds number ${\rm Re}$, which is the ratio of inertial to viscous forces, is also small, we have ${\rm Re}\ll{\rm Ma}\ll1$.
\newline

{\it Perspective.}
The balance between robustness and behavioral variation underlies the evolvability of biological systems. Highly canalized processes \cite{waddington1942canalization}  function despite large genetic and environmental perturbations, allowing them to accumulate genetic variation. These genetic variations represent pre-adaptations, which can be selected for if large environmental changes occur \cite{rutherford1998hsp90, Eshel1998, flatt2005evolutionary}.

The ciliary beat of {\it Chlamydomonas} is known to be functionally robust against mutations in axonemal dyneins in the sense that beating still occurs, though with altered beat amplitude, frequency and shape (see references in the {\it Introduction}). Our study extends these findings by showing that the variation in shape associated with dynein mutations, as well as with other mutations and environmental perturbations, can be accounted for by variation in the curvature sensitivity of the motors. Indeed, beating can still occur when the sign of $\beta'$ changes and when the amplitude of $\beta''$ changes three-fold. We propose that this tolerance of beating to the curvature sensitivity of the motors may have been permissive for the evolution of the molecular heterogeneity of the axonemal dynein family: axonemes contain two or three outer-arm dyneins and several inner-arm dyneins \cite{wickstead2007dyneins}.

An outstanding question is why all extant ciliated organisms share this common family of axonemal dyneins, which suggests that heterogeneity was “locked in” around one billion years ago when last eukaryotic common ancestor lived \cite{mitchell2017evolution}. Additional arguments apply to other components, where there is structural heterogeneity among different species. For example, the radial spokes and central pair are missing in some cilia, additional asymmetrically localized molecules lead to planar or asymmetric beats \cite{dutcher2020asymmetries}, and even the number of doublets can vary from as few as three  \cite{prensier1980motile} to as many as hundreds \cite{Mooseker1978}. Thus, we propose that the functional robustness of the mechanochemistry of sliding filaments tolerates the molecular and structural heterogeneity of the axoneme.
\newline

\begin{acknowledgments}

{\it Acknowledgements.}


\end{acknowledgments}

\section{Materials and Methods}
\subsection{Experiments}
{\it Preparation and reactivation of axonemes.} Axonemes from \textit{Chlamydomonas reinhardtii }cells (received from
chlamy.org) were purified and reactivated. The procedures described in the following are detailed in \cite{alper2013reconstitution}.

Chemicals were purchased from Sigma Aldrich, MO if not stated otherwise. In brief, cells were grown in TAP+P medium under conditions of continious illumination (2x75 W, fluorescent bulb) and air bubbling at 24$^{\circ}$C over the course of 2 days, to a final density of $10^6$ cells/ml. Cilia were isolated using dibucaine, then purified on a 25\% sucrose cushion and demembranated in HMDEK (30mM HEPES-KOH, 5 mM MgSO4, 1 mM DTT, 1 mM EGTA, 50 mM potassium acetate, pH 7.4) augmented with 1\% (v/v) Igpal and 0.2 mM Pefabloc SC. The membrane-free axonemes were resuspended in HMDEK plus 1\% (w/v) polyethylene glycol (molecular weight 20kDa), 30\% sucrose, 0.2 mM Pefabloc and stored at -80$^{\circ}$C. Prior to reactivation, axonemes were thawed at room temperature, then kept on ice. Thawed axonemes were used for up to 2hr.

Reactivation was performed in flow chambers with a depth of 100 $\mu$m, built from easy-cleaned glass and double-sided sticky tape. Thawed axonemes were diluted in HMDEKP reactivation buffer. If not stated otherwise a standard reactivation buffer, containing 1 mM  ATP and an ATP-regeneration system (5 units/ml creatine kinase, 6mM creatine phosphate) was used to maintain a constant ATP concentration. The axoneme dilution was infused into a glass chamber, which was blocked using casein solution (from bovine milk, 2 mg/mL) for 10 min and then sealed with vacuum grease. Prior to imaging, the sample was equilibrated on the microscope for 5 min and data was collected for a maximum time of 20 min.
\newline

{\it Special reactivation conditions.} Temperature series: The temperature was controlled using an objective heater from (Bioptech). If not stated otherwise, the sample temperature was kept constant at 24$^{\circ}$C. Temperature series were acquired by increasing the temperature in 2$^{\circ}$C steps and letting the system equilibrate for 10 min. After equilibration, the target temperature was checked using an inbuilt reference Thermistor. ATP series: The standard buffer (without ATP) was augmented with different amounts of ATP (50, 66, 100, 240, 370, 500, 750, 1000$\,\mu$M). Viscosity: The standard buffer was augmented with Ficol 400 (1\%, 5\%, 10\% (w/v)), then axonemes were added to this solution. Calcium: We used a Ca$^{\rm 2+}$ buffered reactivation solution with a concentration of free Ca$^{\rm 2+}$ of $10^{-5}{\rm M}$ (calculated with the program Maxchelator: https://somapp.ucdmc.ucdavis.edu/pharmacology/bers/ \\ maxchelator). Taxol: The standard buffer was augmented with 10 $\mu{\rm M}$ Taxol, then axonemes were added to this solution
\newline

{\it Imaging of axonemes.} The reactivated axonemes were imaged by phase constrast microscopy, set up on an inverted Zeiss Axiovert S100-TV or Zeiss Observer Z1 microscope using a Zeiss 63× Plan-Apochromat NA 1.4 or a 40x Plan Neofluar NA 1.3 Phase3 oil lens in combination with a 1.6× tube lens and a Zeiss oil condenser (NA 1.4). Movies were acquired using a EoSens 3CL CMOS highspeed camera. The effective pixel size was 139 or 219 nm/pixel. Movies of up to 3000 frames were recorded at a frame rate of 1000 fps.
\subsection{Data analysis}

{\it High precision tracking of isolated axonemes.} To track the shape of the axoneme in each movie frame with nm precision, the Matlab-based software tool FIESTA was used \cite{ruhnow2011tracking}. Prior to tracking, movies were background subtracted to remove static inhomogeneities arising from uneven illumination and dirt particles. The background image contained the mean intensity in each pixel calculated over the entire movie. This procedure increased the signal-to-noise ratio by a factor of 3 \cite{alper2013reconstitution}. Phase-contrast images were inverted. For tracking, a segment size of 733 nm (approximately 5x5 pixels) was used, corresponding to the following program settings: a full width at half maximum of 750 nm, and a ``reduced box size for tracking especially curved filaments'' of 30\%. Along the arc-length of each filament, 25 equally spaced segments were fitted using two-dimensional Gaussian functions.
\newline

{\it Fourier analysis of tangent angle.} Using the x,y positions of the filament shape we calulated the tangent angle $\psi(s,t)$ at every arc-length position in time. To only study the dynamic modes of the waveform we subtracted the static mode (the 0th Fourier mode)  \cite{geyer2016independent} and Fourier-decomposed the tangent angle. The spectum of dynamic modes of the tangent angle shows a peak at the fundamental frequency (first Fourier mode, Fig.~\ref{fig:scheme}D). Because the fundamental mode accounts for $>90$\% of the total power, we neglected the higher harmonics (n = 2,3,4, ...) in all further analysis (see \cite{sartori2016dynamic} for more detail) and only consider the first Fourier mode of $\psi(s,t)$ which is denoted as $\psi(s)$ and will be used hereafter for all dynamic quantities. We calculated the frequency $f=\langle f(s)\rangle $ and the arc-length dependent amplitude $a(s)=|\psi(s)|$ and phase $\varphi(s)=\arg{ \psi(s)}$ profiles of the fundamtal Fourier mode. 
\newline

{\it Polynomial decomposition.} We use shifted Legendre polynomials, $\ell_n(s)$, with normalization $\int_0^1\ell_n(s)\ell_m(s)\d s=\frac{1}{2n+1}\delta_{mn}$, to characterize amplitude and phase profiles. Any function $g(s)$ with support $[0,1]$ can be written as $g(s)=\sum_{n\ge0}g_n\ell_n(s)$, with  $g_n=(2n+1) \int_{0}^1 g(s)\ell_n(s)\d s$. With this definition, $g_0=\int_{0}^1g(s)\d s$ is the mean over the arc-length.
\subsection{Theory}

{\it Mechanical model of the axoneme.} The dynamics of the axoneme is characterized by a balance of hydrodynamic, elastic, and internal sliding forces, produced by motors \cite{machin1958wave, camalet2000generic}. To linear order and in frequency space the force balance equation, which we call Machin equation, is given by
\begin{align}\label{eq:machin}
i\overline{\rm Ma}\psi=\partial_s^4\psi+\partial_s^2f_{\rm s}\quad,
\end{align}
where $f_{\rm s}$ is the dimensionless sliding force density and the a dimensionless constant $\overline{{\rm Ma}}={2\pi f\xi_{\rm n}L^4}/{\kappa}$. A global balance of sliding forces also holds, $\int_0^1f_{\rm s}(s)\d s=F_{\rm b}$, where the basal force is given by $F_{\rm b}=\chi_{\rm b}\Delta_{\rm b}$ with  $\Delta_{\rm b}$ the basal sliding and $\chi_{\rm b}=\chi_{\rm b}'+i\chi_{\rm b}''$ the complex basal response coefficient, with $\chi'_{\rm b},\chi''_{\rm b}>0$.

We take the sliding force density $f_{\rm s}(s)$ generated by motors to depend on the local curvature $\partial_s\psi(s)$ and sliding $\Delta(s)=\Delta_{\rm b}+\psi(s)-\psi(0)$ of doublets \cite{sartori2016dynamic}. In particular,
\begin{align}\label{eq:motmod}
f_{\rm s}=k\Delta+\beta\partial_s\psi\quad,
\end{align}
where $k>0$ is a passive sliding stiffness, and $\beta=\beta'+i\beta''$ is a complex curvature response coefficient. For $\beta''<0$ there is forward (base to tip) wave propagation \cite{sartori2016curvature}. Note that, unlike in \cite{sartori2016curvature}, we allow for instantaneous curvature response via $\beta'$, which is key to characterize amplitude asymmetry. As boundary conditions we use torque and force balances: $\partial_s\psi(0)=F_{\rm b}$, $\partial_s\psi(1)=0$, $\partial_s^2\psi(0)=f_{\rm s}(0)$,  and $\partial_s^2\psi(1)=f_{\rm s}(1)$. Equations \ref{eq:machin} and \ref{eq:motmod} correspond to the solution of a non-linear problem near the point of instability \cite{camalet2000generic}. Note that we have not considered the  role of three dimensional components to the beat, e.g. via twist \cite{sartori2016curvature}.
\newline

{\it Model fitting.} In previous works, we have used the Machin equation and its variants to reproduce the observed beating patterns of cilia \cite{riedel2007molecular,sartori2016curvature}.
For a given value of $\overline{\rm Ma}$, estimated from experiments, and values for the sliding response coefficients $k$, $\beta'$ and $\beta''$, there exists a discrete set of solutions to the system of equations posed by Eqs.~\ref{eq:machin} and \ref{eq:motmod}, the global sliding balance, and the boundary conditions \cite{camalet2000generic, geyer2018computational}. We keep the solution with lowest wavenumber, which is the first one to be excited \cite{sartori2019effect}, so that for a given set of motor response parameters we can determine a unique theoretical waveform, $\{k,\beta',\beta''\}\to\psi_{\rm the}(s)$.

The procedure to fit the model to the data has been extensively described in \cite{geyer2018computational}. In brief, we define a score between the theoretically predicted waveform and the experimentally observed one, $R^2(\psi_{\rm the},\psi_{\rm exp})$, so that $R^2=1$ corresponds to a perfect matching of theory and experiments \cite{sartori2016curvature, geyer2018computational}. We then maximize this score with respect to the three parameters $\{k,\beta',\beta''\}$ using a principal axis algorithm with three different initial points: $\{3 \pi^2, 0, -4 \pi\}$, which corresponds to the plane wave approximation, and two perturbations around it, $\{1.7, -2.1, -5.8\}$ and $\{3 \pi^2, 0.15, -9.6\}$. In addition, to calculate $\overline{\rm Ma}$, we used $\kappa=400{\rm pN}\,\mu {\rm m}^2$ and $\xi_{\rm n}=0.0034 {\rm pN}\,{\rm sec}\,\mu {\rm m}^{-2}$. In the Ficol datasets we adjusted $\xi_{\rm n}$ following our own measurements by factors of $1.1$, $1.6$, and $3.2$ for dilutions of $1\%$, $5\%$ and $10\%$, respectively.
\newline

{\it Low friction limit.} Although Eqs.~\ref{eq:machin} and \ref{eq:motmod} are linear, the boundary conditions result in complicated relationships between the exponential scales in the solution and the motor parameters \cite{geyer2018computational}. We found that such complications are greatly reduced in the limit $\overline{\rm Ma}\to0$. It is particularly convenient to work in terms of curvature $\partial_s\psi$, which can be shown to take the form
\begin{align}
\partial_s\psi=\e^{cs}(\e^{cd(s-1)}-\e^{-cd(s-1)})
\end{align}
with $c=\beta/2$ and $d =\sqrt{1+4k/\beta^2}$. The general condition for the existence of a solution in this limit is simply
\begin{align}\label{eq:tanh}
\tanh(c d) = - \frac{\chi_{\rm b}c d}{\chi_{\rm b}c + \chi}\quad.
\end{align}

As the observed values for the basal stiffness arising from the fits are small, we explore the limit $\chi_{\rm b}\to0$. In this limit Eq.~\ref{eq:tanh} becomes $\tanh(cd)=0$, which implies $cd = in\pi$, with $n=1,2,\ldots$ From this we obtain
\begin{align}\label{eq:c}
\partial_s\psi = \exp\left(\beta's/2\right)\sin\left(n\pi(s-1)\right)\exp\left(i\beta''s/2\right)\quad,
\end{align}
with the sliding stiffness and curvature response coefficients related by $ {\beta}\sqrt{1+{4k}/{\beta^2}}= i2\pi n$. Note that, in the case that $k$ is real, $\beta'=0$ for this limiting case. The dashed line in Fig.~\ref{fig:model} corresponds to the previous equation when $n=1$, which provides very good agreement with the fits for the full model (finite viscosity, non-zero basal compliance, non-zero asymmetry). Note that in Eq.~\ref{eq:c} the amplitude asymmetry is directly related to $\beta'$, with $\beta'<0$ corresponding to decreasing amplitude. Furthermore, the wavelength can be directly read off from this expression, and is given by $\lambda=-4\pi L/\beta''$. This is also in agreement with observations from fitting the full model. 

\bibliography{axo_div}

\clearpage
\onecolumngrid
\section{ Supplementary Material}
\subsection{Variability of shape parameters, within and between axonemes} 

We  analyzed the variability in frequency, amplitude and wavelength within and between axonemes.\
To deterine the within variablity (single axoneme level) we devided time series that contained at least 50 beat cycles into parts of 5 beat cycles. These parts were analyzed by Fourier decomposition and the frequency, mean-amplitude and wavelength of the fundamental mode were extracted \cite{sartori2016dynamic}.

To decide how many parts are nessesary to approximate the variance well we plotted the variance as a function of the number of parts in each time series. In Figure~S1, we performed this analysis on 35 wt axonemes (mean and SEM of the variance is shown).  We find that using 10 parts (50 beat cycles) the variance reaches saturation ( $>80\%$ of the maximum value measured from the longest timesereies of 175 beat cycles). 

\begin{figure}[ht]
\includegraphics[width=1\linewidth]{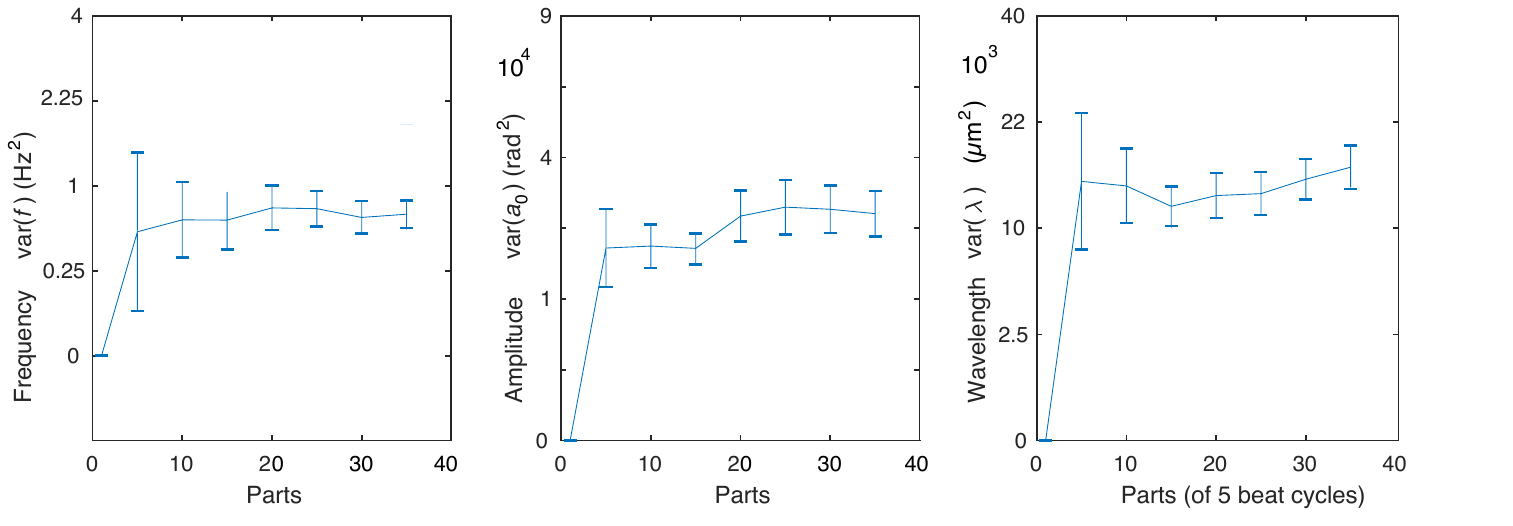}\\[1mm]
{Figure S1. {\it Variance of shape parameters for single axonemes} Variance of the frequency, mean-amplitude and wavelength and their standard deviations for a population of 35 wt axonemes (1mM ATP, 24$^{\circ}$C). On the x-axis the number of analyzed parts (containing 5 beat cycles) is shown.}
\label{fig:SI_para_SD_of_parts}
\end{figure}

The variance analysis was performed on axonemes with more than 50 beat cycles. This allowed us to unambiquiously compare the variation in and between axonemes. The data is summarized in Table S1. We find that the within variance, the variation on the single axoneme level (here, mean of the single axoneme variances) is lower than the in between variance, the variation between axonemes (variance of the mean parameters) for all parameters and in all conditions.  We quantified this for the different shape parameters by the ratios of the between/ within variance, and report the median values over all condition which are approximately 30 for the frequency, 5 for the amplitude and 30 for the wavelength. 





\begin{figure}[ht]

\includegraphics[width=0.85\textwidth]{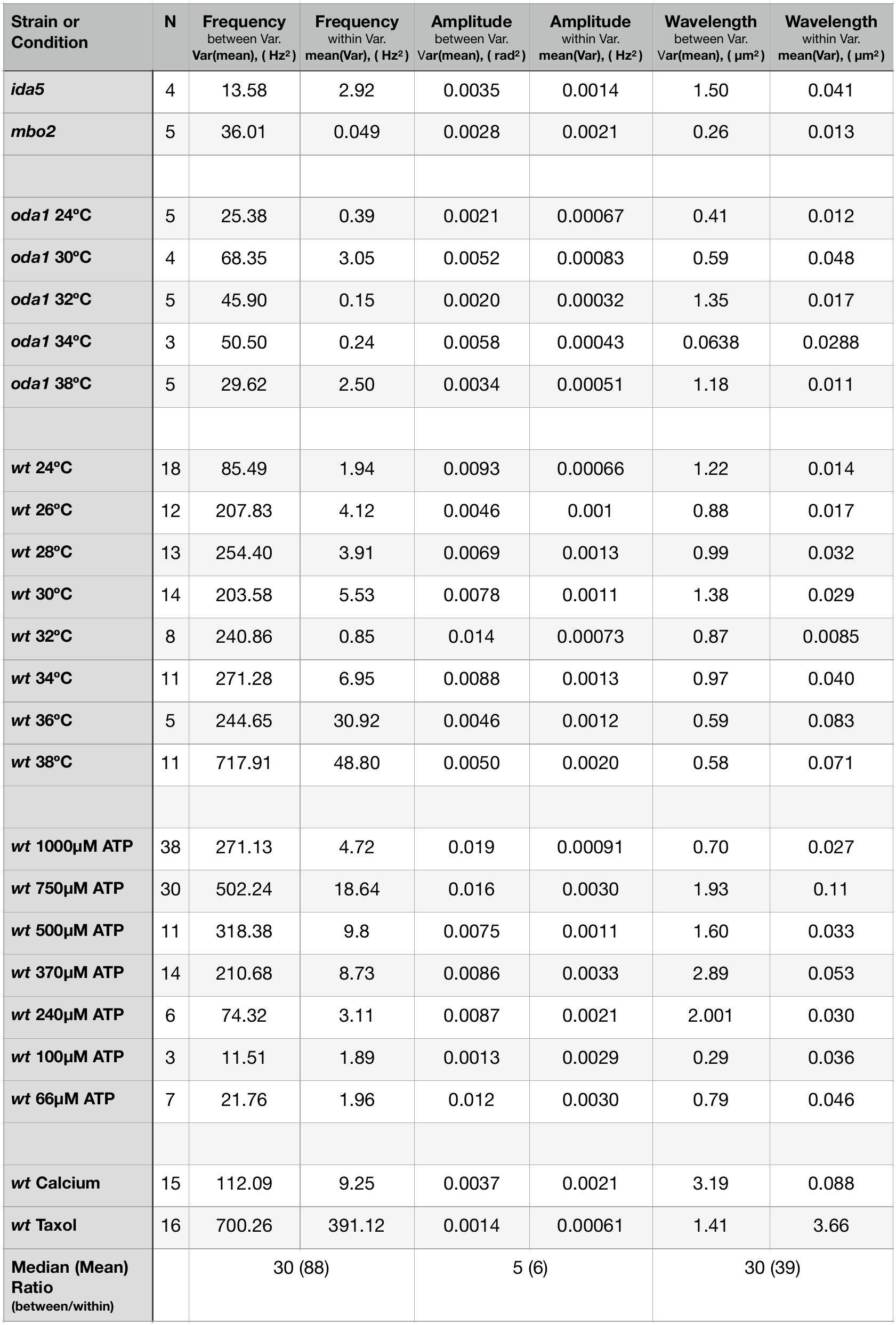}\\[1mm]
{Table S1. {\it Variance analysis of shape parameters} The variances of the frequency, mean-amplitude and wavelength are evaluated for axonemes traces containting 50 beat cycles or more. Shown is the Var(mean), the variance of the population of axonemes within a condition (between variance) and the mean(Var) of single axonemes within one condition (within variance). Mean ratios are the column means, weighted by the sample number.}
	\label{fig:SI_para_single_pop_mut}
\end{figure}

\subsection{Description of appended data file}
The appended file \texttt{analysis\_data.csv} contains the post-processed data used in this work. The first row contains descriptors of each column, whereas the subsequent 498 rows correspond to the axonemes analyzed in this work. The descriptors are
\begin{itemize}
\item \texttt{Labels}: a unique label for each axoneme containing information of the experimental conditions (e.g. \texttt{ATP\_66\_4} corresponds to the fourth axoneme under ATP concentration of $66\,\mu{\rm M}$)
\item \texttt{Length}: the total length $L$ of the axonemes in $\mu{\rm m}$ 
\item \texttt{Frequency}: the frequency $f$ in  ${\rm s}^{-1}$ 
\item \texttt{Amplitude}: the mean amplitude $a_0$
\item \texttt{PcAmplitude1}: the first principal component of the amplitude $a_1^{\rm pc}$
\item \texttt{PcAmplitude2}: the second principal component of the amplitude $a_2^{\rm pc}$
\item \texttt{PcPhase1}: the first principal component of the phase $\varphi_1^{\rm pc}$
\item \texttt{PcPhase2}: the second principal component of the phase $\varphi_2^{\rm pc}$
\item \texttt{LegAmplitude1}: the first Legendre coefficient  of the amplitude $a_1$
\item \texttt{LegAmplitude2}: the second Legendre coefficient  of the amplitude $a_2$
\item \texttt{LegPhase1}: the first Legendre coefficient of the phase $\varphi_1$
\item \texttt{LegPhase2}: the second Legendre coefficient  of the phase $\varphi_2$
\item \texttt{FitScore}: the fitting score $R^2$
\item \texttt{k}: the (dimensionless) shearing stiffness $k$
\item \texttt{Beta'}: the (dimensionless) instantenous curvature response $\beta'$
\item \texttt{Beta''}: the (dimensionless) instantenous curvature response $\beta''$
\end{itemize}
Pre-processed data, along with scripts to perform the analysis in this work, will be made publicly available on-line upon acceptance of the manuscript.





\end{document}